\documentclass[prb,floatfix,reprint,amsmath,amssymb,superscriptaddress,aps]{revtex4-1}
\usepackage{graphicx}
\usepackage{dcolumn}
\usepackage{bm}
\usepackage{color}
\usepackage[T1]{fontenc}
\usepackage{textcomp}
\usepackage{commath}

\begin{document}

\title{Symmetry-resolved magnetoelastoresistance in multivalley bismuth}

\author{Suguru Hosoi}
\email{shosoi@lanl.gov}
\affiliation{Graduate school of Engineering Science, The University of Osaka, Toyonaka, Osaka, 560-8531, Japan}
\affiliation{Los Alamos National Laboratory, Los Alamos, New Mexico, 87545, USA}
\author{Fumu Tachibana}
\affiliation{Graduate school of Engineering Science, The University of Osaka, Toyonaka, Osaka, 560-8531, Japan}
\author{Mai Sakaguchi }
\affiliation{Graduate school of Engineering Science, The University of Osaka, Toyonaka, Osaka, 560-8531, Japan}
\author{Kentaro Ishida}
\affiliation{Graduate school of Engineering Science, The University of Osaka, Toyonaka, Osaka, 560-8531, Japan}
\author{Masaaki Shimozawa}
\affiliation{Graduate school of Engineering Science, The University of Osaka, Toyonaka, Osaka, 560-8531, Japan}
\author{\\Koichi Izawa}
\affiliation{Graduate school of Engineering Science, The University of Osaka, Toyonaka, Osaka, 560-8531, Japan}
\author{Yuki Fuseya}
\affiliation{Department of Engineering Science, University of Electro-Communications, Chofu, Tokyo, 182-8585, Japan}
\author{Yuto Kinoshita}
\affiliation{Institute for Solid State Physics, the University of Tokyo, Kashiwa, Chiba, 277-8581, Japan}
\author{Masashi Tokunaga}
\affiliation{Institute for Solid State Physics, the University of Tokyo, Kashiwa, Chiba, 277-8581, Japan}

\begin{abstract}
We report a symmetry-resolved study of longitudinal magnetoelastoresistance (MER) in the multivalley material bismuth, with the current, uniaxial stress, and magnetic field all applied along the binary axis. The magnitude of MER exhibits a steep increase at low magnetic fields, reaches a peak, and then gradually decreases at higher fields. By decomposing the strain response into symmetric and antisymmetric symmetry channels, we reveal contrasting magnetic field dependencies. Despite the overall non-monotonic field dependence of the MER, the symmetric component remains nearly constant under magnetic fields, suggesting that the valleys in bismuth preserve a rigid-band nature against strain even in the presence of a magnetic field. In contrast, the antisymmetric component, associated with mobility anisotropy, dominates the MER response in a magnetic field. At low magnetic fields, the applied field effectively modifies the apparent mobility of each valley, leading to an enhancement in the magnitude of the antisymmetric MER. At higher fields, field-induced valley polarization further modifies this mobility anisotropy by altering the contributions from each valley’s mobility, accounting for the moderate suppression of the MER. These findings demonstrate that symmetry-resolved MER serves as a powerful probe of valley-dependent electronic states and provides a fundamental platform for understanding the interplay between magnetic field, strain, and charge transport.
\end{abstract}

\maketitle
 
\section{Introduction}
Bismuth has opened up the frontier in solid-state physics by discovering various fascinating phenomena (for a recent review\cite{doi:10.7566/JPSJ.84.012001}): large diamagnetism, high thermoelectric performance\cite{PhysRevLett.98.076603}, high magnetoresistivity\cite{doi:10.1098/rspa.1928.0103,PhysRev.181.1070}, and prominent quantum oscillations in various physical quantities\cite{PhysRevLett.98.166602,PhysRevB.7.602,doi:10.1126/science.1158908}. Underlying these phenomena are the remarkable electronic properties of bismuth, such as ultra-low carrier density, compensated semimetal, very high and anisotropic mobility, large and anisotropic $g$-factor, Dirac electrons, and valley degrees of freedom\cite{doi:10.7566/JPSJ.84.012001}. Among these properties, the valley degrees of freedom\textemdash arising from valleys, which are local energy minima in the conduction band or local maxima in the valence band\textemdash have emerged as one of the most intensively studied topics due to their relevance to the field of valleytronics\cite{SchaibleyNatRevMat2016}. In bismuth, three equivalent electron valleys (e1,e2,e3) and one hole valley are present, and this valley structure is manifested in the characteristic magnetic-field dependence of orbital magnetoresistance (MR), which provides compelling evidence for the presence of valley degrees of freedom\cite{PhysRevX.5.021022,Zhu_2018} and establishes bismuth as an ideal platform for exploring valley-related physics.

An essential technique for realizing valleytronics is the ability to control these valleys. For example, in bismuth, moderate magnetic fields effectively produce finite valley polarization\cite{KulcherNatMater2014,ZhuNatPhys2012,PhysRevB.84.115137}, while even stronger fields (above 40 T) can selectively depopulate one or two valleys depending on the field direction\cite{ZhuNatCommun2017,IwasaSciRep2019}. In addition, uniaxial stress works to lift valley degeneracy\cite{BrandtSovPhys,PhysRevResearch.6.033096}, demonstrating that the valleys in bismuth are highly tunable by both magnetic field and strain. Beyond this tunability, bismuth exhibits a large transport response in both MR\cite{doi:10.1098/rspa.1928.0103,PhysRev.181.1070} and elastoresistance (ER)\cite{PhysRevResearch.6.033096}, making it an exceptionally sensitive platform for quantitatively evaluating the effects of these perturbations. This combination of high valley tunability and pronounced transport sensitivity motivates the key question of how the interplay between magnetic and stress fields governs charge transport in bismuth.

To address this issue, magnetoelastoresistance (MER)—the magnetic-field-dependent counterpart of elastoresistance (ER)—is among the most suitable methods. ER has been established as a powerful probe of anisotropic electronic properties in various quantum materials\cite{doi:10.1126/science.1221713,doi:10.1073/pnas.1605806113,doi:10.1073/pnas.2110501119,RigssNC2015,doi:10.7566/JPSJ.89.064707,PhysRevX.14.031015,PhysRevResearch.2.043293}. Building on this capability, MER offers further insight into how strain modifies transport parameters such as carrier density and mobility, as demonstrated in WTe$_2$\cite{doi:10.1073/pnas.1910695116} and ZrSiSe\cite{JFLinnartznpjQM2024}. Furthermore, the technique becomes even more powerful when symmetry-resolved measurements are performed, which distinguishes between symmetric and antisymmetric strain responses and exploits the fact that strain enables symmetry-selective tuning of material properties. MER is therefore well-suited for probing the magnetic-field evolution in each symmetry channel of bismuth, a material that exhibits exceptionally large ER in both components, particularly at low temperatures\cite{PhysRevResearch.6.033096}.

 In this paper, we present symmetry-resolved MER experiments on bulk bismuth to elucidate how strain and magnetic field affect valley dependent transport. We measure longitudinal MER, where the current, uniaxial stress, and magnetic field are applied along the binary axis. The MER signal of bismuth exhibits non-monotonic magnetic field dependence, reaching a maximum in magnitude at around 0.5 T following the rapid increase at low fields. Beyond this peak at higher fields, clear quantum oscillations emerge and persist up to 80 K—unlike in standard MR, where oscillations vanish at this temperature—highlighting the high sensitivity of MER to the electronic structure. Interestingly, by decomposing MER into each symmetry channel, we find that only the antisymmetric component of MER is sensitive to magnetic fields, whereas its symmetric component remains nearly constant. To explain these contrasting behaviors, we extend a classical transport model based on the ellipsoidal shape of the mobility tensors for each valley. This model, previously demonstrated within the rigid-band approximation to accurately describe the elastoresistance of bismuth at zero magnetic field \cite{PhysRevResearch.6.033096}, is here generalized to incorporate the combined effects of magnetic field and strain. Within this framework, the symmetric ER/MER is independent of mobility, leading to field-insensitive behavior consistent with experimental observations; by contrast, the antisymmetric ER/MER depends on the anisotropy of the mobility tensor and consequently becomes sensitive to magnetic fields, in agreement with experimental observations. Our simple transport model thus captures how magnetic fields modify this anisotropy through two key mechanisms: effective modification of apparent mobility at low fields and field-induced valley polarization at higher fields. This combination of symmetry-resolved MER experiments and transport modeling enables a successful description of charge transport under simultaneous magnetic and stress fields in bismuth and provides complementary insight into how strain and magnetic field control valley degrees of freedom in transport.

\section{Method}
Sample preparations and experimental set-up for transport measurements under uniaxial stress and magnetic field used in this study are described in our previous study \cite{PhysRevResearch.6.033096}. Uniaxial stress was applied to samples attached to the platform \cite{10.1063/5.0008829} using the home-built piezo-driven apparatus based on the design originally reported in Ref. [\onlinecite{10.1063/1.4881611}]. The samples were fabricated to suitable dimensions for ER/MER measurements: typically $\sim 1$ mm (binary:$x$) $\times$ 400 {\textmu}m (bisectrix:$y$) $\times$ 60 {\textmu}m (trigonal:$z$). Four gold wires were attached to the sample surface with silver paste for resistance measurements using the standard four-probe method.
 In order to study the symmetry-resolved response, we have measured the resistance change along the binary directions,
$\Delta R_{xx}(\varepsilon_{ii}) = R_{xx}(\varepsilon_{ii}) - R_{xx}(\varepsilon_{ii}=0)$
($i = x, y$), under two strain configurations, $\varepsilon_{xx}$ and $\varepsilon_{yy}$.
The corresponding ER components are defined as

	\begin{equation}
	{\rm ER}_{\parallel} = \frac{{\rm d} \Delta R_{xx}(\varepsilon_{xx})/R_{xx}(\varepsilon_{xx}=0)}{{\rm d} \varepsilon_{xx}},
	\end{equation}
	
and
	\begin{equation}
	{\rm ER}_{\perp} = \frac{{\rm d} \Delta R_{xx}(\varepsilon_{yy})/R_{xx}(\varepsilon_{yy}=0)}{{\rm d} \varepsilon_{yy}}.
	\end{equation}
In this case, MER is written as ER under a magnetic field, namely, $\mathrm{MER_{\parallel/\perp}} = \mathrm{ER}_{\parallel/\perp}(B)$. To ensure a straightforward interpretation of the MER signal, we restricted the measurements to longitudinal configurations, where the applied current and magnetic fields are parallel to the binary direction in both strain configurations. Because of spatial limitations in the magnet setup, MER measurements for each strain configuration were performed separately using two different superconducting magnets: vertical magnetic fields up to 7 T by solenoid magnet for MER$_{\parallel}$ measurements and horizontal fields up to 4 T by split magnet for MER$_{\perp}$ measurements. Each symmetry-resolved MER is obtained as follows:
\begin{equation}
 \mathrm{MER_{sym}} = \frac{1}{1-\nu_{\rm p}}(\mathrm{ER_{\parallel}}(B) +\mathrm{ER_{\perp}} (B) ),
\end{equation}
\begin{equation}
 \mathrm{MER_{anti}} = \frac{1}{1+\nu_{\rm p}}(\mathrm{ER_{\parallel}}(B) -\mathrm{ER_{\perp}} (B) ),
\end{equation}
where $\nu_{\rm p}$ is an effective Poisson ratio of the platform determined by the experiments ($\nu_{\rm p} \sim - 0.197$). $\mathrm{MER_{sym}}$ and $\mathrm{MER_{anti}}$ represent MER response to the in-plane symmetric strain $\varepsilon_{\rm sym} = (\varepsilon_{xx} + \varepsilon_{yy})/2$ and the symmetry-breaking antisymmetric strain $\varepsilon_{\rm anti} = (\varepsilon_{xx} - \varepsilon_{yy})/2$, respectively.

\begin{figure*}[t]
\centering
\includegraphics[angle=0, width=175mm]{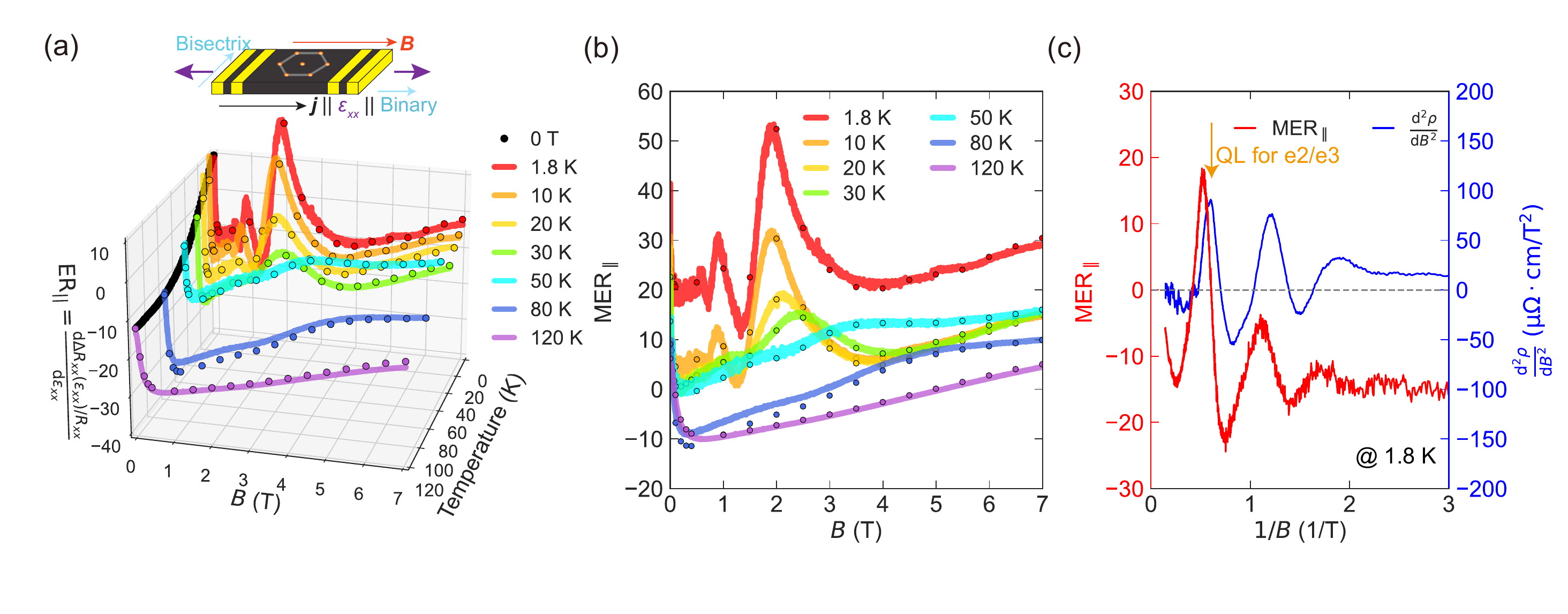}
  \vspace{-10mm}
\caption{MER$_{\parallel}$ study of bismuth. (a)(upper inset) Experimental configuration for MER$_{\parallel}$ measurements. (lower inset) Temperature and magnetic field dependent ER/MER$_{\parallel}$. Circle markers represent data points obtained by strain-sweeping at fixed temperatures and fields, and the line curves represent data obtained by field sweeping at three different strain-fixed conditions. (b)Field-induced changes in MER, $\Delta$MER$_{\parallel} = {\rm ER}_{\parallel}(B)-{\rm ER}_{\parallel} (0)$, at each temperature. The data are vertically shifted for clarity. (c) Comparison of quantum oscillations between MER$_{\parallel}$ (left axis) and MR (right axis) at 1.8 K. The orange arrow represents the quantum limit (QL) for e2/e3 valleys.}
\label{Fig:MER_para}
\end{figure*}

\section{Results and Discussions}
Figure \ref{Fig:MER_para}(a) shows the magnetic field dependence of MER$_{\parallel}$ at various temperatures, along with temperature-dependent ER at 0 T previously reported in Ref. [\onlinecite{PhysRevResearch.6.033096}]. At 120 K, MER$_{\parallel}$ rapidly grows in the negative direction and is subsequently suppressed in higher fields, resulting in the minimum structure below 0.5 T. This overall magnetic field dependence remains essentially unchanged upon cooling, except for the emergence of additional peak structures arising from quantum oscillations, as shown in Fig. \ref{Fig:MER_para}(b). We first discuss the quantum oscillations in MER. Based on the well-defined Landau spectrum of bismuth \cite{PhysRevB.84.115137,Zhu_2018}, the observed prominent oscillations in our experiment are derived from the e2/e3 valleys with a quantum limit of $\sim1.5$ T at the lowest temperature. Interestingly, as shown in Fig. \ref{Fig:MER_para}(c), the largest peak in MER$_{\parallel}$ at 1.8 K lies slightly above this quantum limit, and the peak positions in MER are located near the nodes of the standard MR oscillations. These behaviors originate from the fact that MER is a strain-derivative quantity. To clarify this point, we start from the basic framework that phenomenologically describes quantum oscillations in resistivity $\rho_{\rm osc}$:
\begin{equation}
 \rho_{\rm osc} (B) = \alpha \cos ( 2 \pi F/B + \delta).
\end{equation}
Here, $\alpha$, $F$, and $\delta$ describe the amplitude, frequency, and phase of oscillations, respectively.
The amplitude $\alpha$ roughly relates to the effective mass of the band, and the frequency reflects the cross-sectional area of Fermi surface. The phase $\delta$ is uniquely determined by the band structure. Within the rigid-band approximation underlying this study, strain does not alter the band structure, and $\delta$ therefore remains invariant. In this case, quantum oscillations under strain are given as below:
\begin{equation}
\begin{aligned}
    \frac{{\rm d} \rho_{\rm osc}}{{\rm d} \varepsilon} = &\frac{{\rm d} \alpha}{{\rm d} \varepsilon} \cos  ( 2 \pi F/B + \delta) \\ &+ \frac{2\pi \alpha}{B} \frac{{\rm d} F}{{\rm d} \varepsilon}\cos ( 2 \pi F/B + \delta + \frac{\pi}{2}).
\end{aligned}
\end{equation}
This expression naturally explains the observed phase shift in MER through the second term, whose contribution is determined by the strain-induced change in carrier density. Therefore, the observed large quantum oscillations with a phase shift directly reflect the high strain tunability of valley densities in bismuth, as reported in Ref. [\onlinecite{PhysRevResearch.6.033096}]. This strain tunability also accounts for the robustness of quantum oscillations up to higher temperatures, as shown in Fig. \ref{Fig:MER_para}(b), whereas such signatures are absent in standard MR measurements. Furthermore, we observe a clear shift in the oscillation peaks of MER to higher magnetic field with increasing temperature, in qualitative agreement with previous reports on the temperature evolution of the carrier density in bismuth\cite{J_PMichenaud_1972,PhysRevB.105.235116}. These results collectively demonstrate that MER serves as a highly sensitive probe of the electronic structure.

\begin{figure}[t]
\centering
\includegraphics[angle=0, width=85mm]{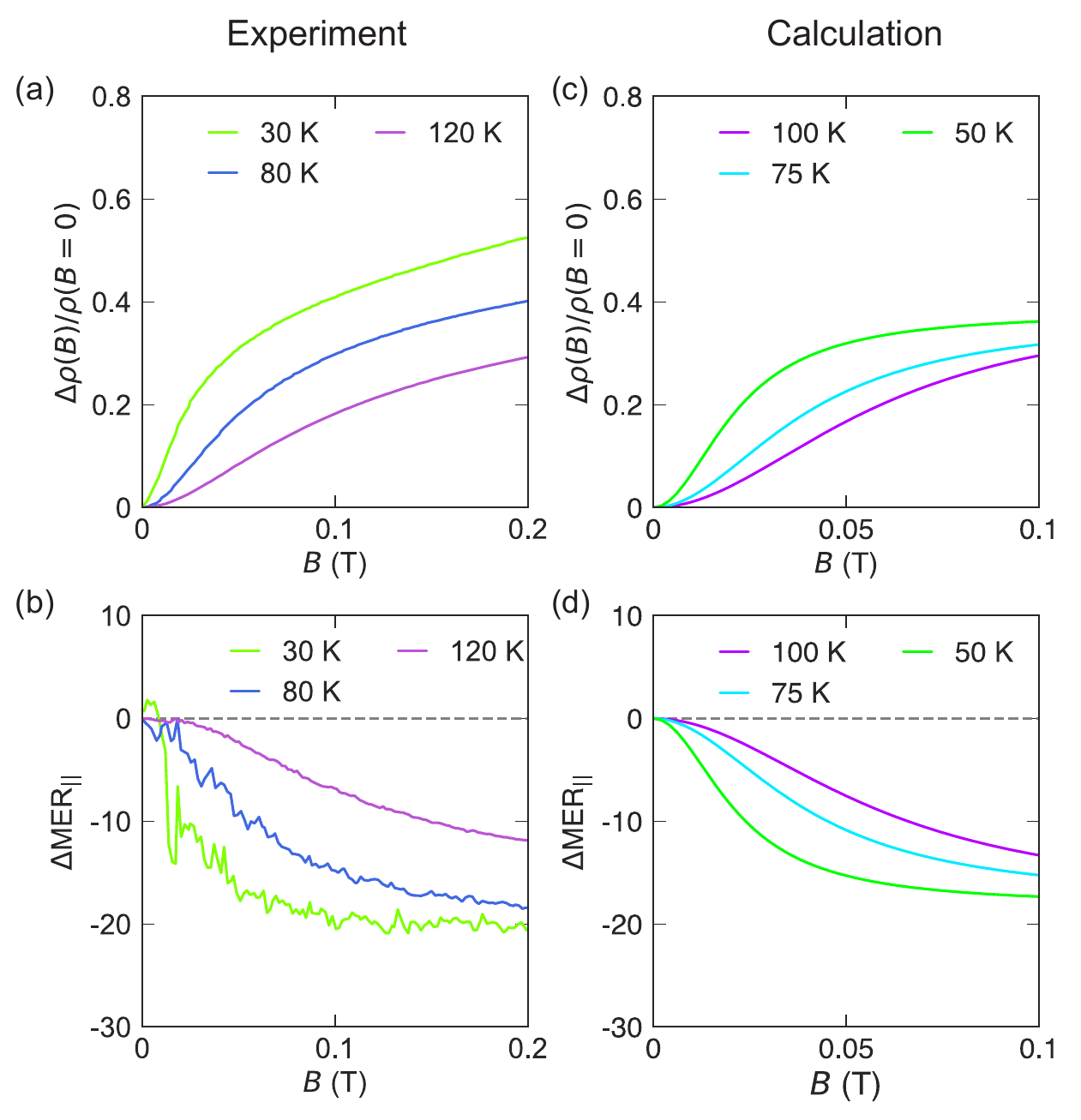}
\caption{Comparison between the low-field behavior of MR and MER  at various temperatures. (a),(b) Experimental results of MR (a) and $\Delta {\rm MER}_{\parallel}$ (b). (c),(d) Numerical calculations of the corresponding quantities using the mobility tensor values from Ref. [\onlinecite{PhysRevX.5.021022}] and valley susceptibility obtained from elasto-quantum oscillation measurements in Ref. [\onlinecite{PhysRevResearch.6.033096}].  }
\label{Fig:MER_vs_MR}
\end{figure}

Next, to address the microscopic mechanism underlying the MER behavior, we compare low-field responses of MR and MER, as shown in Figs. \ref{Fig:MER_vs_MR}(a)(b). Both quantities exhibit a steep initial increase upon application of a magnetic field, followed by a tendency toward saturation. These initial increases become more pronounced at lower temperatures, suggesting a shared origin rooted in field-induced modifications of charge transport. This behavior can be captured within a simple classical transport framework. In fact, the MR of bismuth has been well-explained by using a general expression with mobility tensor and an appropriate magnetic field tensor $\hat{B}$, as introduced by Aubrey \cite{JEAubrey_1971}:
\begin{equation}
	\hat{\sigma} = n e \left[  \hat{\mu_0}^{-1} \pm \hat{B} \right]^{-1},
    \label{eq:sig_B}
\end{equation}
where $n$ is carrier density, $\hat{\mu_0}$ is the mobility tensor, and $e$ is the elementary charge. The sign $\pm$ corresponds to the sign of charge. For multivalley systems like bismuth, total conductivity can be obtained through the summation of conductivity tensors from each valley.
 In this framework, the field evolution of the MR is essentially governed by the  effective mobility, defined as  $\hat{\mu}_{\rm eff} (B) = [\hat{\mu_0}^{-1} + \hat{B}]^{-1}$.
 Using the mobility tensor reported in Ref. [\onlinecite{PhysRevX.5.021022}], we qualitatively reproduce the initial increase of MR, as shown in Fig. \ref{Fig:MER_vs_MR}(c). We note that the numerical values of each mobility tensor component are subject to uncertainties due to their extraction from published graphs; therefore, we focus here on the qualitative behavior of the magnetic field dependence. Importantly, the initial steep rise in MER is naturally reproduced by calculations based on the same mobility tensors, as shown in Fig. \ref{Fig:MER_vs_MR}(d). The correlated trend between the initial increase in MR and MER is also reproduced, with a steeper rise in MR leading to a correspondingly steeper rise in MER. In addition, both MR and MER exhibit increasingly steep rises as the temperature is lowered. These qualitative agreements between experiment and calculation support the conclusion that $\hat{\mu}_{\rm eff}$ accounts for the low-field behavior of MER.

 A symmetry-resolved study provides more complementary insights into the mechanism of MER. Under the crystal symmetry of bismuth, MER response can be decomposed into symmetric (MER$_{\rm sym }$) and antisymmetric (MER$_{\rm anti}$) components. Introducing the concept of strain-valley susceptibility, defined as  $\chi_{\Gamma} = \frac{1}{n} \frac{{\rm d} n}{{\rm d} \varepsilon_{\Gamma}} $, where $n$ is the carrier density and $\varepsilon_{\Gamma}$ is the symmetry-resolved strain, the MER can be expressed using a simple extension of the ER formalism\cite{PhysRevResearch.6.033096} under the rigid-band approximation:
  \begin{equation}
 \mathrm{MER}_{\rm sym} = - \chi_{\rm sym},
 \label{MER_sym_Formula}
\end{equation}
\begin{equation}
\mathrm{MER}_{\rm anti} = \gamma \chi_{\rm anti}.
 \label{MER_ant_Formula}
 \end{equation}
 Here, $\gamma$ denotes the anisotropy of mobility among three electron valleys\cite{PhysRevResearch.6.033096}. Given that previous elasto-quantum oscillation measurements under magnetic field yield values of $\chi_{\rm sym}$ and $\chi_{\rm anti}$ that are comparable in magnitude to those independently determined from zero-field ER studies\cite{PhysRevResearch.6.033096}, the magnetic field dependence of the valley susceptibilities is likely modest. Based on this condition, Eqs. (\ref{MER_sym_Formula})(\ref{MER_ant_Formula}) predict that only MER$_{\rm anti}$ can be sensitive to magnetic field through $\gamma$, while MER$_{\rm sym }$ is field-insensitive.

 It should be noted that the application of an in-plane magnetic field reduces the in-plane rotational symmetry from trigonal to orthorhombic. Such symmetry lowering can, in principle, lead to mixing between the symmetric and antisymmetric strain channels, potentially affecting the validity of the MER$_{\rm sym}$/MER$_{\rm anti}$ decomposition. However, the field-induced orthorhombicity is negligibly small ($\varepsilon_{xx}^{B} \sim 0.001 \%$ at 5 T along the binary direction \cite{IwasaSciRep2019}) compared to the applied perturbative strain ($\Delta \varepsilon_{xx} \sim \pm 0.08\%$ at 120 K), and therefore the symmetry decomposition remains valid under our experimental conditions.
\begin{figure}[t]
\centering
\includegraphics[angle=0, width=80mm]{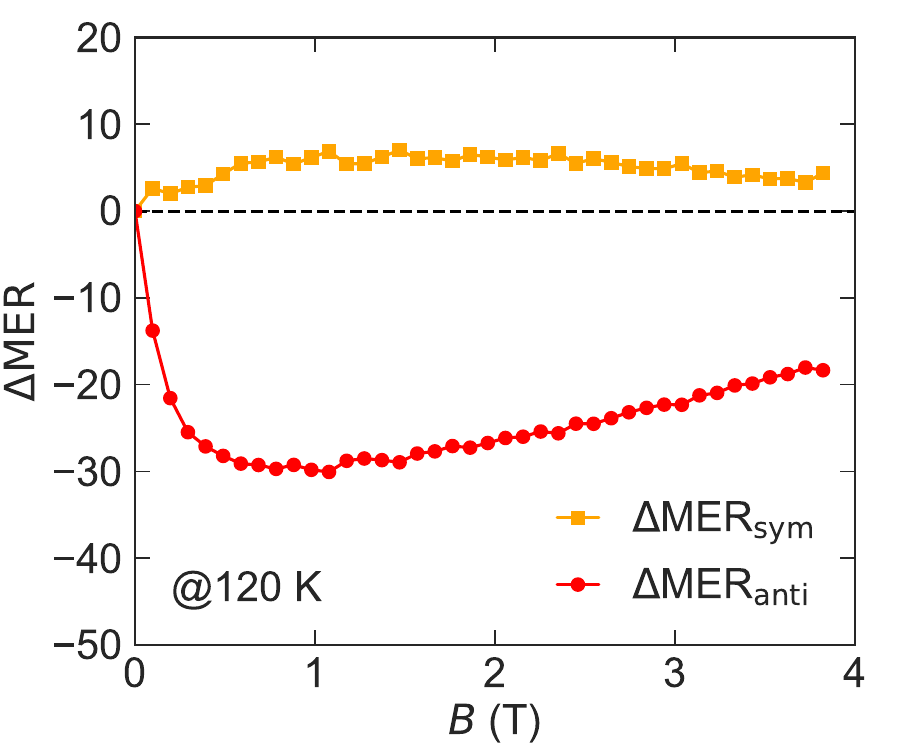}
\caption{Field evolution of symmetry-resolved $\Delta$MER at 120 K. Red circles and orange squares represent antisymmetric and symmetric components of $\Delta$MER, respectively.}
\label{Fig:SYym_MER}
\end{figure}

\begin{figure}[t]
\centering
\includegraphics[angle=0, width=80mm]{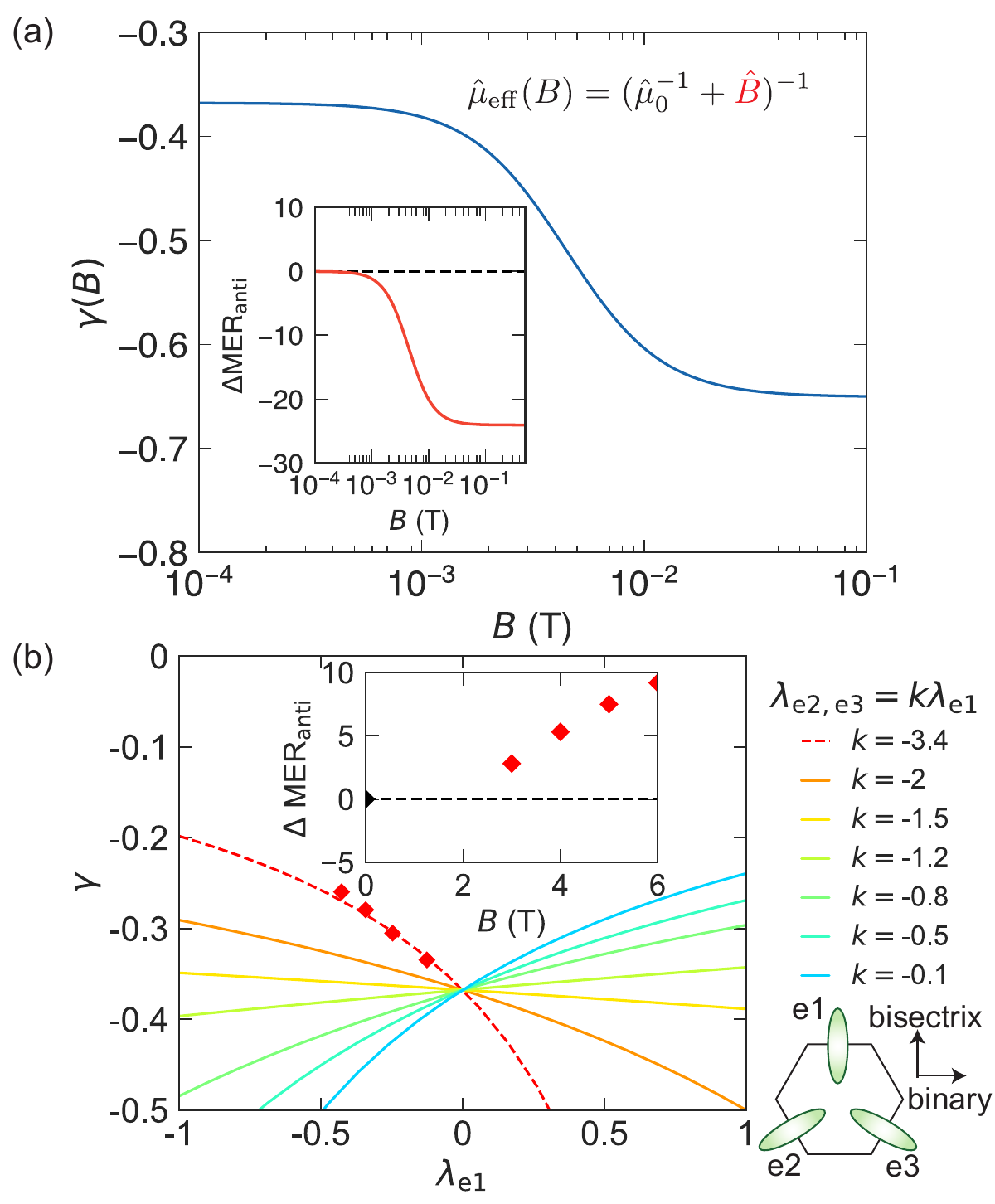}
\caption{Numerical calculations of mobility anisotropy $\gamma$ arising from apparent mobility modification (a) and valley population changes (b). (a) $\gamma$ modified by $\hat{\mu}_{\rm eff}$, evaluated using Eq. (\ref{eq:sig_B}) and the mobility tensor from Ref. [\onlinecite{PhysRevX.5.021022}], with magnetic field applied along the binary axis. The inset shows the corresponding $\Delta {\rm MER_{anti}}$ by using valley susceptibility $\chi_{\rm anti} \sim 85$ determined from zero-field ER measurements\cite{PhysRevResearch.6.033096}. (b) $\gamma$ as a function of the deviation in carrier density of e1 valley $\lambda_{\rm e1}$. The carrier densities of e2,e3 valleys are defined by $\lambda_{\rm e2,e3} = k \lambda_{\rm e1}$ for various values of $k$. The red diamonds show $\gamma$ values obtained from the carrier density calculated from Landau quantized energy dispersion\cite{PhysRevB.84.115137}, which are well described by $k = -3.4$ (dashed curve). The inset displays the corresponding relative changes in $\Delta {\rm MER_{\rm anti}}$ at several magnetic fields. For simplicity, we neglect magnetic field dependence of mobility $\mu$ in this calculation to capture essential effects of valley polarization. The right panel shows the assignment of the electron valleys (e1, e2, e3).}
\label{Fig:Calc_MER}
\end{figure}

Figure \ref{Fig:SYym_MER} shows the symmetry-resolved field dependence of $\Delta \mathrm{MER} = \mathrm{MER}(B) - \mathrm{MER}(0)$ in bismuth at 120 K, where quantum oscillation effects are absent. As expected from Eqs. (\ref{MER_sym_Formula})(\ref{MER_ant_Formula}), $\Delta$MER$_{\rm sym}$ exhibits only minor changes; this suggests that $\chi_{\rm sym}$ is essentially independent of magnetic field, validating the rigid band assumption with respect to strain even under magnetic fields. Meanwhile, $\Delta$MER$_{\rm anti}$ rapidly increases in the negative direction, which is consistent with the calculated magnetic field dependence of $\gamma$ based on $\hat{\mu}_{\rm eff} (B)$, as shown in Fig. \ref{Fig:Calc_MER}(a). This agreement demonstrates the validity of this framework for describing valley transport under both magnetic field and strain, and confirms that Aubrey’s magnetic field correction term successfully accounts for $\Delta$MER$_{\rm anti}$ in the weak-field region through the modulation of valley-dependent mobility anisotropy.

In the higher magnetic field regime, some extension to the current model is required. The slight decreases in $\Delta$MER$_{\rm anti}$ above 1 T cannot be explained by $\hat{\mu}_{\rm eff}(B)$ if one assumes that that original mobility $\hat{\mu_0}$ is unchanged from its zero magnetic field values. One possible explanation is a field-induced modification of $\hat{\mu}_0$, but it is difficult to quantify its effect since no established model exists for field-modified mobility that fully accounts for the magnetic field dependence of standard MR\cite{ZhuNatCommun2017}. However, the angular dependence of MR in bismuth can still be well explained without invoking such modifications\cite{PhysRevB.103.125148}. For example, in thin-film bismuth under rotating in-plane fields, large anisotropic MR (AMR) and planar Hall exhibit higher-order oscillations at low fields, which reduce to predominantly two-fold symmetry at higher fields and low temperatures\cite{PhysRevResearch.2.022029}. According to Ref. [{\onlinecite{PhysRevB.103.125148}}], such drastic changes in angular dependence are attributed to field-induced changes in carrier density. Therefore, these classical transport models for MR still remain applicable in the present magnetic field regime, not by requiring significant modifications to the mobility $\hat{\mu}_0$, but incorporating field-induced changes in carrier density. 

 Building on this understanding, a field-induced valley polarization offers another plausible mechanism that modifies the carrier density in each valley, thereby accounting for the field dependence of MER$_{\rm anti}$. Because $\gamma$ reflects the mobility anisotropy among three electron valleys, an imbalance of valley density can also effectively modify $\gamma$ by changing the relative contributions from each valley to total transport. For instance, $\gamma$ would be effectively compensated if the population of the higher mobility e1 valley along the binary direction decreases while that of the lower mobility e2/e3 valleys increases, leading to the suppression of $\Delta$MER$_{\rm anti}$, as observed. Indeed, field-induced valley polarization in bismuth is well established, as evidenced by its large magnetostriction\cite{PhysRevB.26.2552,KulcherNatMater2014,IwasaSciRep2019}. In particular, this type of valley polarization ($\Delta n_{\rm pol} = n_{\rm e1} - n_{\rm e2,e3} <0$) is known to occur under magnetic fields along the binary direction.

 To evaluate its effect on $\Delta$MER$_{\rm anti}$, we calculated how a valley polarization modifies the anisotropy parameter $\gamma$. For simplicity, we only considered the carrier density term, incorporating both stain-induced and field-induced changes:
 \begin{eqnarray}
 n_{\rm e1} (\varepsilon_{\rm anti}) &=& n(1 + \chi_{\rm anti} \varepsilon_{\rm anti}  + \lambda_{\rm e1})\\
 n_{\rm e2,e3} (\varepsilon_{\rm anti}) &=& n(1 - \chi_{\rm anti} \varepsilon_{\rm anti}/2+ \lambda_{\rm e2,e3})\\
 n_{\rm h} (\varepsilon_{\rm anti}) &=& n(3 + \lambda_{\rm e1} + 2\lambda_{\rm e2,e3})
 \end{eqnarray}
 Here, $\lambda_{\rm e1}$ and $\lambda_{\rm e2,e3}$ represent offset contributions due to magnetic field, defined relative to the original carrier density $n$ for electron valley. Charge neutrality was assumed for determining the hole carrier density $n_{\rm h}$. To reduce the number of free parameters, we introduced a proportionality constant $k$ such that $\lambda_{\rm e2,e3} = k \lambda_{\rm e1}$, and examined the $\gamma$ values under various valley polarization conditions with different $k$ values as shown in Fig. \ref{Fig:Calc_MER}(b). To capture the anisotropic changes induced by magnetic fields along the binary direction, we focus on the case of $k<0$. The impact of $\lambda_{\rm e1}$ on $\gamma$ strongly depends on $k$: for small $|k|<1$, a decrease in e1 valley population ($\lambda_{\rm e1}<0$) increases the anisotropy $\gamma$, while for $|k|>1.2$, it tends to suppress $\gamma$. To estimate the actual change in valley populations, we used the Landau-quantized energy dispersion model from Ref. [\onlinecite{PhysRevB.84.115137}]. The calculated $\gamma$ values at various fields are plotted as markers in Fig. \ref{Fig:Calc_MER}(b), and they correspond well with the case of $k = -3.4$. This change results in a positive shift of $\Delta{\rm MER}_{\rm anti}$, as shown in the insets, which is consistent with our experimental observations. Thus, by combining the effects of $\hat{\mu}_{\rm eff}(B)$ and $\Delta n_{\rm pol}$, the model qualitatively reproduces the overall magnetic field dependence of $\Delta{\rm MER}_{\rm anti}$ (see Appendix for details). These findings suggest that mobility anisotropy parameter $\gamma$ plays an essential role in determining the magnetic field dependence of ${\rm MER_{anti}}$.

\begin{figure*}[t]
\centering
\includegraphics[angle=0, width=165mm]{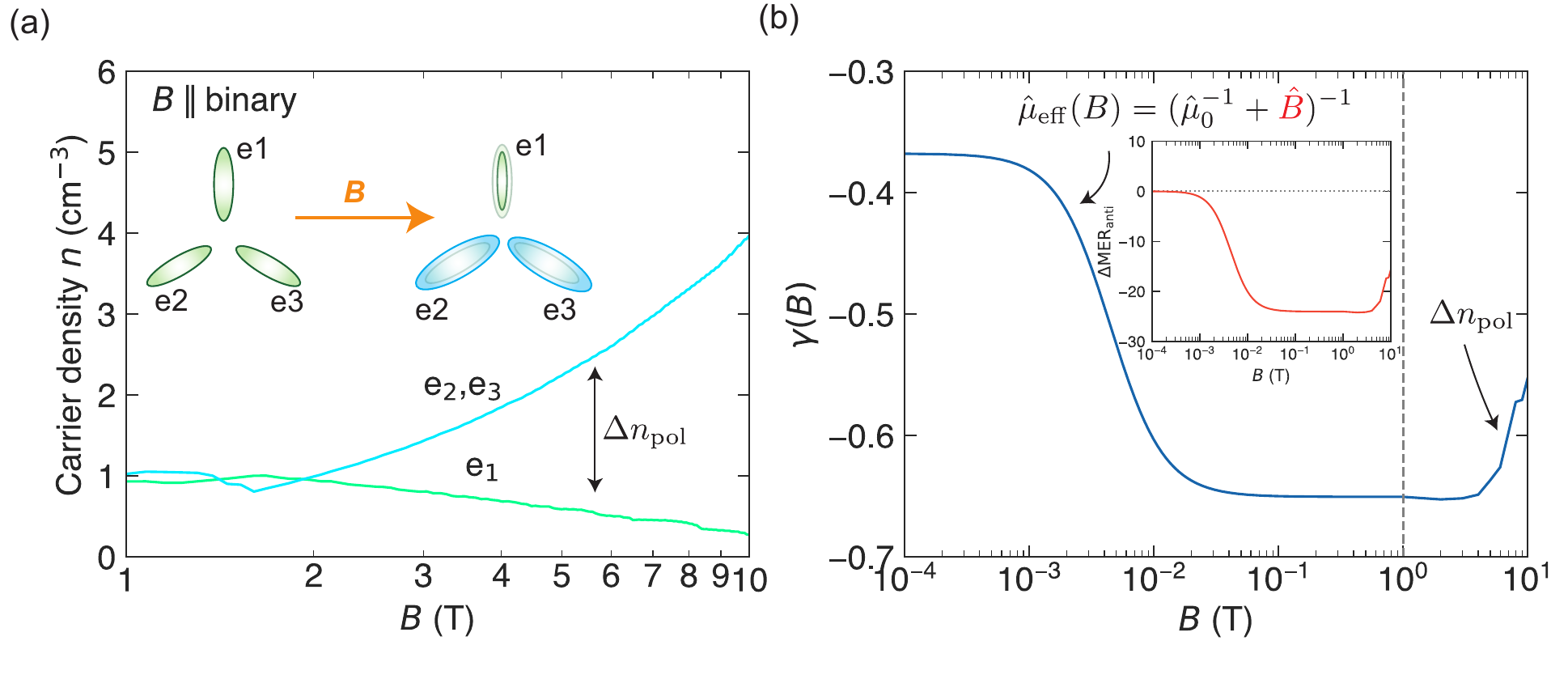}
\caption{Numerical calculations of  valley populations (a) and mobility anisotropy $\gamma$ (b). (a) Calculated magnetic-field dependence of valley populations, reproduced from Ref. [\onlinecite{PhysRevB.84.115137}]. The inset shows a schematic picture of field-induced valley polarizations under $B \parallel $ binary. (b) $\gamma$, evaluated using Eq. (\ref{eq:sig_B}) and the mobility tensor from Ref. [\onlinecite{PhysRevX.5.021022}], with magnetic field applied along the binary axis. For $B > 1$~T (vertical dashed line), the effect of valley polarization is incorporated.  The inset shows the corresponding $\Delta {\rm MER_{\rm anti}}$ using valley susceptibility $\chi_{\rm anti} = 85$.}
\label{Fig:MER_fullmodel}
\end{figure*}

 \section{Conclusion}
  In summary, we have performed symmetry-resolved magnetoelastoresistance (MER) measurements in bismuth and uncovered distinct symmetry-dependent responses. MER$_{\rm anti}$ shows non-monotonic magnetic field dependence, while MER$_{\rm sym}$ remains nearly unchanged. These results can be well explained by the simple classical transport theory within the rigid band approximation. The origin of field-dependent MER lies in the anisotropy of the mobility tensor among valleys. In the low-field region, the application of a magnetic field effectively modifies valley mobilities, leading to an initial rapid increase of MER$_{\rm anti}$. In higher fields, a field-induced finite valley polarization leads to the reduction in MER$_{\rm anti}$ through changing the contribution from each valley's mobility. Our results provide a general framework for understanding valley transport under the combined effects of magnetic field and strain.

	\section*{Acknowledgements}
We thank N. Miura for the provided bismuth samples. This work was supported by Grants-in-Aid for Scientific Research (KAKENHI) (Nos. JP18H01167, JP20K20901, JP22H01939, JP22K03522, JP22K18690, JP23H00268, JP23H04862, JP23K17879) and Grand-in-Aid for Scientific Research on innovative areas ``Quantum Liquid Crystals'' (No. JP20H05162) from Japan Society for the Promotion of Science. S.H. and M. Shimozawa were supported by the Multidisciplinary Research Laboratory System (MRL), The University of Osaka. S.H. was supported by Toyota Riken Scholar Program. S.H. is currently supported by the Director's Funded Postdoctoral Fellowship through the Laboratory Directed Research and Development Program at Los Alamos National Laboratory.

\section*{APPENDIX}

 The valley populations under magnetic field were recalculated using the model introduced in Ref. [\onlinecite{PhysRevB.84.115137}]. The electron valleys are described by the extended Dirac model with additional $g$-factor.
 \begin{equation}
    E_{n,\pm} = \sqrt{\Delta^2 + 2 \Delta (n + \frac{1}{2} \pm \frac{1}{2}) \frac{m_{\rm e}}{m_{\rm c}} \beta_0 B + \frac{\hbar^2 k_z^2}{2 m_z}} \pm \frac{g'}{2} \frac{\beta_0}{2} B,
 \end{equation}
where $\beta_0 = |e|\hbar / m_{\rm e} c$, $n$ is the Landau level index, and the $\pm$ sign corresponds to spin.  
$\Delta$ is half of the band gap at the $L$ point, $m_c$ is the cyclotron mass of electrons, and $m_e$ is the free electron mass.  $g'$ is the additional $g$-factor, and $k_z$ and $m_z$ are the wavevector and effective mass along the field direction, respectively. On the other hand, the hole dispersion at $T$ point is described by a nonrelativistic approximation:

 \begin{equation}
    E_0 + \Delta - E = (n + \frac{1}{2} ) \hbar \Omega_{\rm c} + \frac{\hbar^2 k_z^2}{2 M_z} \pm \frac{G}{2} \mu_{\rm B} B,
 \end{equation}
where $\mu_{\rm B}$ is the Bohr magneton, and $E_0$ is the hybridization energy between electron and hole valleys. $\Omega_{c}$, $M_{\rm c}$, and $G$ represent cyclotron frequency, effective mass, and $g$-factor for holes, respectively. Field-dependent carrier density for electrons is reproduced as shown in Fig. \ref{Fig:MER_fullmodel}(a). All parameter values were taken from Ref.~[\onlinecite{PhysRevB.84.115137}]. By applying magnetic field along the binary direction, valley polarization $\Delta n_{\rm pol} <0 $ is enhanced.

 In the main text, we separately discussed two mechanisms that describe the field dependence of MER$_{\rm anti}$: effective mobility $\hat{\mu}_{\rm eff} (B)$ and valley polarization $\Delta n_{\rm pol}$. Here, Fig. \ref{Fig:MER_fullmodel}(b) shows the complementary calculation results of $\gamma (B)$, incorporating both effects simultaneously. Since the valley density under magnetic fields is based on the model that describes the Landau quantized spectrum, the effective field range is $1 < B < 10$ T, and hence field-modified carrier density changes are included above 1 T. The model qualitatively reproduces the magnetic-field-dependent trend of ${\rm MER_{anti}}$, which exhibits the broad minimum structure around 1 T.

\end{document}